\documentclass[12pt]{article}

\newcommand{\articleonly}[1]{#1}
\newcommand{\lncsonly}[1]{}

\articleonly{\usepackage{fullpage}}
\usepackage{url}
\usepackage{xspace} 
\usepackage{alltt}
\usepackage[T1]{fontenc}
\usepackage{amssymb}
\usepackage{breakcites}

\newcommand{\Hex}[1]{\hspace{#1ex}}
\newcommand{\Vex}[1]{\vspace{#1ex}}
\newenvironment{code}{\Vex{0.5}\begin{alltt}\small}{\end{alltt}\Vex{0.5}}
\newcommand{\eqnnum}[2]{}
\newcommand\co[1]{\mbox{\tt\small #1}} %
\newcommand\m[1]{\mbox{$#1$}} %
\newcommand\p[1]{\m{#1}}
\newcommand{\defn}[1]{\textit{#1}} %
\def\mathify#1{\ifmmode{\mbox{$#1$}}\else\mbox{$#1$}\fi}

\newcommand{\cotr}{\co{\m{\!}.T}\xspace}
\newcommand{\cof}{\co{\m{\!}.F}\xspace}
\newcommand{\cou}{\co{\m{\!}.U}\xspace}
\newcommand{\cocs}{\co{\m{\!}.CS}\xspace}
\newcommand\IF{\mathify{\leftarrow}\xspace}
\newcommand\AND{\mathify{\land}\xspace}
\newcommand\NOT{\mathify{\neg}\xspace}
\newcommand\OR{\mathify{\lor}\xspace}

\newcommand\SOME{\mathify{\exists}\xspace}
\newcommand\EACH{\mathify{\forall}\xspace}
\newcommand\IN{\mathify{\in}\xspace}
\newcommand\T{\mathify{T}\xspace}
\newcommand\F{\mathify{F}\xspace}
\newcommand\UD{\mathify{U}\xspace}

\newcommand{\mypar}[1]{\Vex{-1}\paragraph{\lncsonly{}#1.\,}}
\newenvironment{myexample}{\Vex{.5}\par\textbf{\textit{Example}}.\,}{\hfill$\blacksquare$\par}

\usepackage{enumitem}
\setlist{leftmargin=4ex, itemsep=.6ex}

\begin{document}

\newcommand\ack{
This work was supported in part by NSF under grants 
  CCF-1414078, %
  CCF-1954837, %
  CNS-1421893, %
  and IIS-1447549, %
  and ONR under grant N00014-20-1-2751. %
}

\title{Knowledge of Uncertain Worlds:\\ Programming with Logical Constraints\articleonly{\thanks{\ack
    }}}
\articleonly{\author{Yanhong A. Liu \Hex{6} Scott D. Stoller\\Computer Science Department, Stony Brook University\\liu@cs.stonybrook.edu~~ stoller@cs.stonybrook.edu}
\date{}}
\lncsonly{}

\maketitle

\begin{abstract}
Programming with logic for sophisticated applications must deal with
recursion and negation, which together have created significant challenges
in logic, leading to many different, conflicting semantics of rules.  This
paper describes a unified language, DA logic, for design and analysis
logic, based on the unifying founded semantics and constraint semantics,
that supports the power and ease of programming with different intended
semantics.  The key idea is to provide meta-constraints, support the use of
uncertain information in the form of either undefined values or possible
combinations of values or both, and promote the use of knowledge units that can be instantiated by any new predicates, including predicates with additional arguments.
\Vex{-1} \lncsonly{}
\end{abstract}

\section{Introduction}
\label{sec-intro}

Programming with logic has allowed many design and analysis problems to be
expressed more easily and clearly at a high level.  Examples include
problems in program analysis, network management, security frameworks, and
decision support~\cite{liu18LPappl-wbook}.
However, when sophisticated %
problems require reasoning with negation and recursion, possibly causing
contradiction in cyclic reasoning, programming with logic has been a
challenge.  Many languages and semantics have been proposed,
e.g.,~\cite{fitting85,GelLif88,van+91well},
but they have
different underlying assumptions that are conflicting and subtle%
, and each is suitable for only certain kinds of problems.

This paper describes a unified language, DA logic, for design and analysis
logic, for programming with logic using logical constraints.  It supports
logic rules with unrestricted negation in recursion, as well as
unrestricted universal and existential quantification.  It is based on the
unifying founded semantics and constraint
semantics~\cite{LiuSto18Founded-LFCS,LiuSto20Founded-JLC}, and it supports
the power and ease of programming with different intended semantics without
causing contradictions in cyclic reasoning.
\begin{itemize}

\item 
The language provides meta-constraints on %
predicates.  These meta-\lncsonly{}constraints capture the different underlying assumptions of different logic language semantics. %
\item 
The language supports the use of uncertain information in the results of
different semantics, in the form of either undefined values or possible
combinations of values~or~both.
\item
The language further supports the use of knowledge units that can be instantiated by any new predicates, including predicates with additional arguments.

\end{itemize}

Together, the language allows complex problems to be expressed clearly and easily, where different assumptions can be easily used, combined, and compared for expressing and solving a problem modularly, unit by unit.

We present examples for different games that show the power and ease of
programming with DA logic.
We use games because interdependent winning and losing positions taken by
competing players naturally give rise to negation in recursion.
We also discuss and describe support for
restricted parameters and recursive uses of knowledge units.

The rest of the paper is organized as follows.
Section~\ref{sec-motiv} discusses the need of easier programming with logic
when faced with negation in recursion.
Section~\ref{sec-lang} describes the unified language, DA logic.
Section~\ref{sec-sem} presents the formal definition of the semantics of DA
logic, as well as its consistency, correctness, and decidability.
Section~\ref{sec-examp} develops additional examples for different games.
Section~\ref{sec-disc} explains restricted parameters and recursive uses of
knowledge units.
Section~\ref{sec-related} discusses related work and concludes.

This paper is a revised and extended version of Liu and
Stoller~\cite{LiuSto20LogicalConstraints-LFCS}.  The revisions include
many expanded explanations to make the paper more self-contained and easier to read, as well as general improvements throughout.  The extension is mainly the new Section~\ref{sec-disc} on
support for restricted parameters and recursive uses of knowledge units.

\section{Need of easier programming with logic}
\label{sec-motiv}

We discuss the challenges of %
programming with negation
and recursion and the need of easier programming with logic.
We explain the basic ideas of well-known previous language semantics as well as
founded semantics and constraint semantics, and give an overview of the proposed
solutions.
We use a small well-known example, the win-not-win game,
for illustration.

\mypar{Win-not-win game}
Given a set of moves for a game, consider the following rule, called the win rule. It says that \co{x} is a
winning position if there is a move from \co{x} to \co{y} and \co{y} is not
a winning position.
\begin{code}
 win(x) \IF move(x,y) \AND \NOT win(y) \eqnnum{onerule}{1}
\end{code}

This seems to be a reasonable rule, because, besides giving the conditions
for \co{x} to be a winning position, it also suggests
that, if there is no move from \co{x}, then \co{x} is a losing position,
and if \co{x} is neither a winning nor a losing position, then \co{x} is a draw position.
This captures the rule for winning and losing for many games,
including in chess for the King to not be captured, giving winning, losing,
and draw positions.

However, there could be problems.  For example if there is a \co{move(1,1)}
for some position~\co{1}, then the win rule would give \co{win(1) \IF \NOT
  win(1)}, and thus the truth value of \co{win(1)} becomes unclear.

\mypar{Inductive definitions}
Instead of the single win rule, one could use the following three rules to
determine the winning, losing, and draw positions.
\begin{code}
 win(x) \IF \SOME y | move(x,y) \AND lose(y)
 lose(x) \IF \EACH y | \NOT move(x,y) \OR win(y)
 draw(x) \IF \NOT win(x) \AND \NOT lose(x) \eqnnum{threerule}{3}
\end{code}

The first two rules are used~\cite{hou2010fo,dasseville2015semantics} to form inductive definitions~\cite{moschovakis1974elementary}.
avoiding the potential problems of the single win rule.  The base case is
the set of positions that have no moves to any other position and thus are
losing positions based on the second rule.

With winning and losing positions defined, the draw positions are the
remaining positions, which are those in cycles of moves that have no moves
to losing positions.

These three rules spell out the intended meaning of winning, losing, and
draw as implied by the single win rule. However, clearly, these rules are
much more cumbersome than the single win rule.

\mypar{Well-founded semantics}
Indeed, with well-founded semantics (WFS)~\cite{van+91well}, which computes
a 3-valued model, the single win rule above gives \co{win(x)} being true,
false, or unknown for each \co{x}, corresponding exactly to \co{x} being a
winning, losing, or draw position, respectively. 

WFS is highly non-trivial---informally, it is defined by the least fixed
point of a transformation that combines what is usually
called the one-step derivability operator, \m{T_p}, and the element-wise negation of
the operator \m{U_p} for computing what is called the
greated unfounded set, yielding a single 3-valued model~\cite{van+91well};
it involves computing an alternating
fixed point %
or an iterated fixed point. %

However, \co{win(x)} being 3-valued in WFS does not allow the three
outcomes to be used as three predicates or sets for further computation;
the three predicates defined by the three rules do allow this.

For example, there is no way to use draw positions (that is, positions for
which \co{win} is unknown) explicitly, say to find all reachable nodes
following another kind of moves from draw positions.  One might try to do
this by adding the following two additional rules to the single win rule:
\begin{code}
 lose(x) \IF \NOT win(x)
 draw(x) \IF \NOT win(x) \AND \NOT lose(x)
\end{code}
However, the result
is that \co{draw(x)} is false for all positions for which \co{win(x)} is
true or false, and \co{draw(x)} is unknown for all draw positions.

\mypar{Stable model semantics} 
Stable model semantics (SMS)~\cite{GelLif88} computes a set of 2-valued
models, instead of a single 3-valued model.  It has been used for solving
many constraint problems in answer set programming (ASP), because its set
of 2-valued models can provide the set of satisfying solutions.

SMS is also highly non-trivial---informally, it is defined by guessing a
truth assignment, expanding each rule into all possible instances,
computing what is called the reduct by deleting rules whose negated
conditions cannot be satisfied and deleting negated conditions in remaining
rules, and then computing a minimum model of the resulting rules, yielding
one model in a set of 2-valued models~\cite{GelLif88}; in general, the
number of guesses and resulting models can be exponential.

For the single win rule, if besides some winning and losing positions,
there is a separate cycle of even length, say \co{move(1,2)} and
\co{move(2,1)}, then the win rule would give \co{win(1) \IF \NOT win(2)}
and \co{win(2) \IF \NOT win(1)}.
Instead of \co{win} being unknown for positions \co{1} and \co{2} as in
WFS, SMS returns two models: one with \co{win} being true for \co{1} and
false for \co{2}, and one with \co{win} being true for \co{2} and false for
\co{1}.  This is a very different interpretation of the win rule.

For the single win rule above, when there are draw positions, SMS may also
return just the empty set, that is, the set with no models at all.  For
example, if besides some winning and losing positions, there is a separate
cycle of moves of odd length, say simply \co{move(1,1)}, then SMS returns
just the empty set.  This is clearly not the desired semantics for the
win-not-win game.

\mypar{Founded semantics and constraint semantics}
Founded semantics and constraint semantics~\cite{LiuSto18Founded-LFCS,LiuSto20Founded-JLC} unify
different prior semantics.  They define a 3-valued model and a set 
of 2-valued models, respectively.
They allow different underlying assumptions to
be specified for each predicate.
Specifically:
\begin{enumerate}

\item Each predicate can be declared \defn{certain} (that is, everything
  about the predicate being true (\T) are given or can be inferred by
  following the rules, and the rest are false (\F)) or \defn{uncertain}
  (that is, everything about the predicate being \T or \F are given or can
  be inferred, and the rest are undefined (\UD)), except a predicate must be
  uncertain if it depends on negation in recursion or on uncertain
  predicates.

\item Each uncertain predicate can be further declared \defn{complete}
  (that is, all rules with the predicate in the conclusion are given, and
  thus before inferring \T and \F, completion rules can be added to
  define the negation of the predicate using the negation of the conditions
  of those given rules) or not, except a predicate must be not complete if
  it depends on predicates that are uncertain and not complete.

\item Each uncertain and complete predicate can be declared \defn{closed}
  (that is, an assertion of the predicate is made \F, called self-false, if
  inferring it to be \T requires assuming itself to be \T) or not.
  Being closed is needed to match WFS and SMS theoretically, but is not
  needed to give the desired meaning for any example we found in previous literature.

\end{enumerate}
Founded semantics infers \T, \F, and \UD using a simple least fixed point,
with additionally computing self-false assertions for closed predicates, 
if any, in each iteration.
Constraint semantics then extends everything \UD to be combinations of \T
and \F that satisfy everything given
as constraints.

For the win-not-win game, one can write the single win rule,
with the default assumption that \co{win} is complete, that is, the win
rule is the only rule that infers \co{win}, which is an implicit assumption
underlying WFS and SMS.
\begin{itemize}
\item With founded semantics, the three rules that use inductive
  definitions can be automatically derived, and true, false, and undefined
  positions for \co{win} are inferred, corresponding to the three
  predicates from inductive definitions and the 3-valued results from WFS.

\item Then constraint semantics, if desired, computes all combinations of
  true and false values for the undefined values for the draw positions,
  that satisfy all the rules as constraints.  It equals SMS for the single
  win rule.
\end{itemize}

Explicit declaration in founded semantics and constraint semantics makes programming and understanding much easier.  For example,
in WFS and SMS, if nothing is said about some \co{p}, then
\co{p} is false.  When this is not desired, some programming tricks are
used to get around it.  For example, with SMS, to allow \co{p} to be
possibly true in some models, one could introduce some new \co{q} and two
new rules, \co{p \IF \NOT q} and \co{q \IF \NOT p}, to make it possible
that, in some models, \co{p} is true and \co{q} is false.  Founded
semantics and constraint semantics allow \co{p} to be explicitly declared
uncertain and not complete.

Founded semantics and constraint semantics also allow unrestricted
universal and existential quantifications and unrestricted nesting of
Boolean operators; these are not supported in WFS and SMS.

However, founded semantics and constraint semantics alone do not address
how to use different semantics seamlessly in a single logic program.

\mypar{Programming with logical constraints}

Because different assumptions and semantics help solve different problems
or different parts of a problem, easier programming with logic requires
supporting all assumptions and semantics in a simple and integrated design.

This paper treats different assumptions as different meta-constraints for
expressing a problem or parts of a problem, and support results from
different semantics to be used easily and directly.  For the win-not-win
game:
\begin{itemize}

\item The positions for which \co{win} is true, false, and undefined in
  founded semantics are captured using three automatically derived
  predicates, \co{win.T}, \co{win.F}, and \co{win.U}, respectively,
  corresponding exactly to the inductively defined \co{win}, \co{lose}, and
  \co{draw}, respectively.  These predicates can be used explicitly and
  directly for further reasoning, unlike with the truth values in WFS or
  founded semantics.

\item The constraint semantics of the given rule for \co{win} and facts for \co{move} is captured
  using an automatically derived predicate \co{CS}.  For a model \co{m} in
  the constraint semantics, \co{CS(m)} is true, also denoted as \co{m \IN
    CS}, and we use \co{m.win(x)} to denote the truth value of \co{win(x)}
  in model \co{m}.  Predicate \co{CS} can be used directly for further
  reasoning, unlike the set of models in SMS or constraint semantics.
\end{itemize}

More fundamentally, we must enable easy specification of problems with
reusable parts and where different parts may use different assumptions and
semantics.  To that end, we introduce knowledge units.  DA logic supports
instantiation and re-use of existing units, and allows predicates in any
existing units to be bound to other given predicates, including predicates
with additional arguments.

Even with all this power, DA logic is decidable, because it does not include function symbols and is over finite domains.

Table~\ref{tab:summary} summarizes the meta-constraints that can be used to
express different assumptions, corresponding declarations and resulting
predicates in founded semantics and constraint semantics, and corresponding
other prior semantics if all predicates use the same meta-constraint.
Columns 2 and 4 are presented and proved in our prior
work~\cite{LiuSto20Founded-JLC}.  Columns 1 and 3 are introduced in DA
logic:
\begin{itemize}
\item Each meta-constraint in column 1 specifies the corresponding declarations in column 2.  For example, \co{complete(\p{P})} specifies that \p{P} is declared uncertain, complete, and not closed.
Note that the four meta-constraints capture all possible
combinations of declarations.

\item In column 3, \p{P}\cotr, \p{P}\cof, and \p{P}\cou are predicates that are true for a tuple of arguments if and only if \p{P} is \T, \F, and \UD, respectively, for that tuple of arguments in founded semantics.
\p{K} denotes a knowledge unit, and \p{K\cocs} denotes the constraint semantics of \p{K};
for a model \p{m} in \p{K\cocs}, \p{m}\co{.}\p{P} is a predicate that 
has the same truth values as predicate \p{P} in model \p{m}.

\end{itemize}
These will be described precisely in Sections~\ref{sec-lang} and~\ref{sec-sem}.  
\newcommand{\tabularm}[2]{\mbox{\Hex{-1}}\begin{tabular}[t]{@{~}p{24ex}@{}}#1\\\hline#2\end{tabular}}
\begin{table}[t]
    \centering
    \begin{tabular}{@{}l@{~}|p{20ex}|p{21ex}|p{25ex}@{\Hex{-1}}}\hline
Meta-constraint & \multicolumn{2}{c|}{Founded/Constraint Semantics} 
                                    & Other Prior Semantics\\
\cline{2-3}
on Predicate \p{P}  & Declarations on \p{P} & Resulting Predicates &\\
\hline\hline
\co{certain(\p{P})}	& certain				& \p{P}\cotr, \p{P}\cof	
                                    & Stratified (Perfect,\newline Inductive Definition)\\
\hline
\co{open(\p{P})}	& uncertain,\newline not complete	& \tabularm{\p{P}\cotr, \p{P}\cof, \p{P}\cou}{\p{m}\co{.}\p{P} for \p{m} \IN \p{K}\cocs} 
                                    & \tabularm{\mbox{~}}{First-Order Logic}\\
\hline
\co{complete(\p{P})}	& uncertain,\newline complete, not closed\Hex{-2}	& as above
                                    & \tabularm{Fitting (Kripke-Kleene)}{Supported}\\
\hline
\co{closed(\p{P})}	& uncertain,\newline complete, closed	& as above 	
                                    & \tabularm{WFS}{SMS}\\
\hline
    \end{tabular}
    \caption{Meta-constraints and corresponding prior semantics.}
    \label{tab:summary}
\end{table}

\section{DA logic}
\label{sec-lang}

This section presents the syntax and informal meaning of DA logic, for
design and analysis logic.  The rule form described under ``Conjunctive
rules with unrestricted negation'' is the same as the core language in our
prior work on founded semantics and constraint semantics, for which we gave
a precise semantics~\cite{LiuSto18Founded-LFCS,LiuSto20Founded-JLC}.  Disjunction and
quantification are mentioned as extensions in our prior
work~\cite{LiuSto18Founded-LFCS,LiuSto20Founded-JLC}.  The other features are new.

\mypar{Knowledge unit}

A \defn{program} is a set of knowledge units.  A \defn{knowledge unit},
abbreviated as \defn{kunit}, is a set of rules, facts, and
meta-constraints, defined below.  The definition of a kunit has the
following form, where $K$ is the name of the kunit, and \p{body} is a set
of rules, facts, meta-constraints, and instantiations of other kunits:
\begin{code}
 kunit \p{K}:
   \p{body}
\end{code}
The scope of a predicate is the kunit in which it appears.  Predicates with the same name, but appearing in different kunits, are distinct.
\begin{myexample}
A kunit for the single win rule is
\begin{code}
 kunit win\_unit:
   win(x) \IF move(x,y) \AND \NOT win(y)
\end{code}\Vex{-3}
\end{myexample}

Kunits provide structure and allow knowledge to be re-used in other contexts by instantiation, as described below.  

\mypar{Conjunctive rules with unrestricted negation}

We first present a simple core form of logic rules and then describe additional constructs that can appear in rules.  The core form of a rule is the following, where any \m{P_i} may be preceded with \NOT:
\begin{equation}
  \label{eqn-rule}
  Q(X_1, ... ,X_a) ~\IF~ P_1(X_{11}, ... ,X_{1a_1}) ~\AND~ ... ~\AND~ P_h(X_{h1}, ... ,X_{ha_h})
\end{equation}
Symbols \IF, \AND, and \NOT indicate backward implication, conjunction, and
negation, respectively.  \m{h} is a natural number.  Each \m{P_i}
(respectively \m{Q}) is a predicate of finite number \m{a_i} (respectively
\m{a}) of arguments.
Each argument \m{X_k} and \m{X_{ij}} is a constant or a variable, and each
variable in the arguments of \m{Q} must also be in the arguments of some
\m{P_i}.  In arguments of predicates in example programs, we use numbers
for constants and letters for variables.

If \m{h = 0}, there is no \m{P_i} or \m{X_{ij}}, and each \m{X_k} must be a
constant, in which case \m{Q(X_1, ..., X_a)} is called a \defn{fact}.  For
the rest of the paper, ``rule'' refers only to the case where \m{h \geq 1},
in which case the left side of the backward implication is called the
\defn{conclusion}, the right side is called the \defn{body}, and each
conjunct in the body is called a \defn{hypothesis}.

These rules have the same syntax as in Datalog with negation, but are used
here in a more general setting, because variables can range over complex
values, such as constraint models, as described below.

\mypar{Predicates as sets}

We use a syntactic sugar in which a predicate $P$ is also regarded as the
set of $x$ such that $P(x)$ holds.  For example, we may write \co{move =
  \{(1,2), (1,3)\}} instead of the facts \co{move(1,2)} and \co{move(1,3)};
to ensure the equality holds, this shorthand is used only when there are no
other facts or rules defining the predicate.

\mypar{Disjunction}

The hypotheses of a rule may be combined using disjunction as well as conjunction.  Conjunction and disjunction may be nested arbitrarily.

\mypar{Quantification}

Existential and universal quantifications in the hypotheses of rules are written using the following notations:
\begin{equation}
  \label{eq:quant}
\begin{tabular}[c]{@{}ll@{}}
  $\SOME~ X_1,~...,~X_n~\co{|}~ Y$ & ~~existential quantification\\
  $\EACH~ X_1,~...,~X_n~\co{|}~ Y$ & ~~universal quantification
\end{tabular}
\end{equation}
In quantifications of this form, the domain of each quantified variable
\m{X_k} is the set of all constants in the containing kunit.

As syntactic sugar, a domain can be specified for a quantified variable,
using a unary predicate regarded as a set.  For example, \co{\SOME x \IN
  win | move(x,x)} is syntactic sugar for \co{\SOME x | win(x) \AND
  move(x,x)}, and \co{\EACH x in win | move(x,x)} is syntactic sugar for
\linebreak
\co{\EACH x | \NOT win(x) \OR move(x,x)}.

\mypar{Meta-constraints}

Assumptions about predicates are indicated in programs using the
meta-constraints in column~1 of Table~\ref{tab:summary}.  Each
meta-constraint specifies the declarations listed in column~2 of
Table~\ref{tab:summary}.  For example, if a kunit contains
\co{open(\p{P})}, we say that $P$ is declared uncertain and not complete in
that kunit.  In each kunit, exactly one meta-constraint must be given for
each predicate.

Meta-constraint \co{certain(\p{P})} means that each 
assertion of \p{P} has a unique true (\T) or false (\F) value.
Meta-constraint \co{uncertain(\p{P})} means that each assertion of \p{P}
has a unique true, false, or undefined (\UD) value. 
Meta-constraint \co{complete(\p{P})} means that all rules with \p{P} in the
conclusion are given in the containing kunit.  
Meta-constraint \co{closed(\p{P})} means that an assertion of \p{P} is made
false, called \defn{self-false}, if inferring it to be true using the given
rules and facts requires assuming itself to be true.

A predicate in the conclusion of a rule is said to be \defn{defined} using
the predicates or their negation in the hypotheses of the rule, and this
defined-ness relation is transitive.
If a predicate \p{P} is not defined transitively using its own negation and
is not defined transitively using a predicate that is defined transitively
using its own negation,
then it is given the meta-constraint \co{certain(\p{P})} by default.
Otherwise, 
it is given \co{complete(\p{P})} by default.

\mypar{Using kunits with instantiation}

The body of a kunit $K_2$ can use another kunit $K$ using an instantiation of the form:
\begin{equation}
\co{use}~K~(P_1=Q_1(Y_{1,1}, ..., Y_{1,b_1}),\, ...,\, P_n=Q_n(Y_{n,1}, ..., Y_{n,b_n}))
\end{equation}
By definition, this has the effect of applying the following substitution to the body
of $K$ and inlining the result in the body of $K_2$: for each $i$ in
$1..n$, replace each occurrence $P_i(X_1,...,X_a)$ of predicate $P_i$ with
$Q_i(X_1,...,X_a,Y_{i,1}, ..., Y_{i,b_i})$.  Note that arguments of $Q_i$
specified in the \co{use} construct are appended to the argument list of
each occurrence of $P_i$ in $K$, hence the number of such arguments must be
$\mbox{arity}(Q_i)-\mbox{arity}(P_i)$.  
When $P_i$ and $Q_i$ have the same arity, we simply write $P_i=Q_i$ in the
\co{use} construct.

The determination of default meta-constraints, and the check for having
exactly one meta-constraint per predicate, are performed after expansion of
all \co{use} constructs.

A kunit $K_2$ has a \defn{use-dependency} on kunit $K$ if $K_2$ uses $K$.  The use-dependency relation must be acyclic.
We have not found intrinsically good reasons for uses to be cyclic.  However, there is no real difficulty in supporting circular uses.  Section~\ref{sec-disc} discusses pros and cons of circular uses and extensions to support circular uses.

\begin{myexample}
For the example kunit \co{win\_unit} given earlier in this section, the following kunit is an instantiation of the win-not-win game with different predicates for moving and winning:
\begin{code}
 kunit win2\_unit:
   use win\_unit (move = move2, win = win2)
\end{code}\Vex{-3}
\end{myexample}

In many logic languages, including our prior work on founded semantics
\cite{LiuSto18Founded-LFCS,LiuSto20Founded-JLC}, a program is an unstructured set of rules and
facts.  The structure and re-use provided by kunits is vital for
expressing knowledge modularly, building large conceptual models, and developing large practical applications.

\mypar{Referencing founded semantics}

The founded semantics of a predicate $P$ in a kunit $K$
(formally defined in Section \ref{subsec:semantics}) can be referenced in $K$ using special
predicates $P\cotr$, $P\cof$, and $P\cou$, one for each of the three truth
values \T, \F, and \UD.  For each truth value $t$,
$P.t(c_1,...,c_a)$ is true if $P(c_1,...,c_a)$ has truth value $t$, and is
false otherwise.

Note that the founded semantics of $K$ can be referenced in another kunit $K_2$ by simply adding \co{use $K$}, i.e., instantiating $K$ in $K_2$ without replacing any predicate, when predicates of $K$ and $K_2$ are
disjoint.  Otherwise, an instantiation with predicate replacements can be used to avoid name collisions.
Our could also add a language feature for referencing the founded semantics of $K$ using predicates of the form $K.P\cotr$, $K.P\cof$, and $K.P\cou$, instead of using instantiation.

To ensure that the semantics of $P$ is fully determined before these
predicates are used, 
$P$ cannot be defined transitively using these predicates.
Predicates that reference founded semantics are implicitly given
the meta-constraint \co{certain} and can appear only in rule bodies.

When referencing the undefined part of a predicate, it is sometimes
desirable to prune uninteresting values.  For example, consider the rule
\co{draw(x)} \co{\IF win.U(x)}.  If the kunit contains constants
representing players as well as positions, \co{win(\p{X})} is undefined
when $X$ is a player, and the user wants \co{draw} to hold only for
positions, then the user could add to the rule 
a conjunct \co{move(x,y){\,}\OR{\,}move(y,x)}, to select \co{x} that are
positions in moves.

\mypar{Referencing constraint semantics}

The constraint semantics of a kunit $K$ (formally defined in Section \ref{subsec:semantics}) can be referenced in another kunit
$K_2$ using the special predicate $K\cocs$.
Using this special predicate in any rule in $K_2$ has the effect of adding
each constraint model of $K$ as an element in the domain (that is, set of
constants) of $K_2$.  In other words, the possible values of variables in
$K_2$ include the constraint models of $K$.  The assertion $K\cocs(m)$ is
true when $m$ is a constraint model of $K$ and is false for all other
constants.

Note that the constraint semantics of $K$ cannot be referenced from within $K$; this ensures that the set of constraint models is fully defined before it is referenced.
The constraint semantics of $K$ cannot be referenced by instantiating $K$;
this is why we need to introduce a language feature for referencing it using predicates of the form $K\cocs$.

The constraint models of a kunit $K$ can be referenced using $K\cocs$ only
if $K$ does not reference its own founded semantics (using predicates such
as $P\cou$).  
This restriction is needed to prevent constraint models from containing contradictions such as the following:
suppose \co{P.U(0)} is true in the founded model of a kunit \co{K}, 
and \co{K} has at least one constraint model \co{m};
then \co{P.U(0)} must also be true in \co{m},
but \co{P(0)} must be true or false, not undefined, 
in \co{m}, because \co{m} is 2-valued.
A kunit $K_2$ has a \defn{CS-dependency} on another kunit $K$ if $K_2$ uses
$K\cocs$.  The CS-dependency relation must be acyclic.

When the value of a variable $X$ is a constraint model of $K$, a predicate
$P$ of $K$ can be referenced using the notation $X.P$.  If the value of $X$
is not a constraint model, or $P$ is not a predicate defined in that
constraint model, then $X.P$ is undefined for all arguments.

Predicates that reference constraint semantics are implicitly given
the meta-constraint \co{certain} and can appear only in rule bodies.

\section{Formal definition of semantics of DA logic}
\label{sec-sem}

\articleonly{
\newcounter{thmcounter}
\newenvironment{theorem}{
\refstepcounter{thmcounter}
\medskip\noindent\textbf{\textit{Theorem \thethmcounter}}.}{\hfill}
}
\newenvironment{myproof}{\articleonly{\Vex{1}}\lncsonly{\Vex{-1}}{\bf Proof:\,}}{\hfill$\blacksquare$}

\newcommand{\union}{\cup}

\newcommand{\pgm}{\pi}
\newcommand{\comb}{{\it Combine}}
\newcommand{\addinv}{{\it AddInv}}
\newcommand{\cmpl}{{\it Cmpl}}
\newcommand{\lfp}{{\it LFP}}
\newcommand{\lfpscc}{{\it LFPbySCC}}
\newcommand{\addneg}{{\it AddNeg}}

\newcommand{\founded}{{\it Founded}}
\newcommand{\constraint}{{\it Constraint}}

\newcommand{\dg}{{\it DG}}
\newcommand{\nameneg}{{\it NameNeg}}
\newcommand{\proj}[2]{{\it Proj}(#1,#2)}

\newcommand{\selffalse}{{\it SelfFalse}}

This section extends the definitions of founded semantics and constraint
semantics in \cite{LiuSto18Founded-LFCS,LiuSto20Founded-JLC} to handle the new features of DA logic.

Handling kunits is relatively straightforward.  Because each kunit defines
a distinct set of predicates, the founded semantics of the program is
simply a collection of the founded semantics of its kunits, and similarly
for the constraint semantics.  All \co{use} constructs in a kunit are
expanded, as described in Section~\ref{sec-lang}, before considering its
semantics.  Therefore, the constants, facts, rules, and meta-constraints of
a kunit include the corresponding elements (appropriately instantiated) of
the kunits it uses.

Handling references to founded semantics and constraint semantics requires changes in the definitions of domain, literal, interpretation, and dependency graph.

Handling disjunction, which is mentioned as an extension
in~\cite{LiuSto18Founded-LFCS,LiuSto20Founded-JLC}
but not considered in the detailed
definitions, requires changes in the definition of completion rules and the
handling of closed predicates.

The paragraphs ``Founded semantics of DA logic without closed declarations'', ``Least fixed point'', and ``Constraint semantics of DA logic'' are essentially the same as in \cite{LiuSto18Founded-LFCS,LiuSto20Founded-JLC}; they are included for completeness.

When we say that a predicate is certain, complete, or closed, we mean that it has that declaration in column 2 of Table~\ref{tab:summary} from its  meta-constraint.

\subsection{Preliminary definitions}
\Vex{1}
\mypar{Atoms, literals, and projection}
Let $\pgm$ be a program.  Let $K$ be a kunit in $\pgm$.  A predicate is {\em intensional} in $K$ if it appears in the conclusion of at least one rule in $K$; otherwise, it is {\em extensional} in $K$.  
The {\em domain} of $K$ is the union of the following sets: the set of constants in $K$, and for each kunit $K_1$ such that $K_1\cocs$ appears in $K$, the set of constraint models of $K_1$.  Constraint models are formally defined in the last paragraph of Section \ref{subsec:semantics}.  The requirement that the CS-dependency relation is acyclic ensures the constraint models of $K_1$ are determined before the semantics of $K$ is considered.

An {\em atom} of $K$ is a formula $P(c_1,...,c_a)$ formed by applying a predicate $P$ in $K$ with arity $a$ to $a$ constants in the domain of $K$.
A {\em literal} of $K$ is a formula of the form $P(c_1,...,c_a)$ or $P\cof(c_1,...,c_a)$, for any atom $P(c_1,...,c_a)$ of $K$ where $P$ is a predicate that does not reference founded semantics or constraint semantics.
These are called {\em positive literals} and {\em negative literals} for  $P(c_1,...,c_a)$, respectively.  
A set of literals is {\em consistent} if it does not contain positive and negative literals for the same atom. The {\em projection} of a kunit $K$ onto a set $S$ of predicates, denoted $\proj{K}{S}$, contains all facts of $K$ for predicates in $S$ and all rules of $K$ whose conclusions contain predicates in $S$.

\mypar{Interpretations, ground instances, models, and derivability}
An {\em interpretation} $I$ of $K$ is a consistent set of literals of $K$.
Interpretations are generally 3-valued. 
\begin{itemize}

\item For a predicate \m{P} that does not reference founded or constraint
  semantics, $P(c_1,...,c_a)$ is true (\T) in $I$ if $I$ contains
  $P(c_1,...,c_a)$, is false (\F) in $I$ if $I$ contains
  $P\cof(c_1,...,c_a)$, and is undefined (\UD) in $I$ if $I$ contains
  neither $P(c_1,...,c_a)$ nor $P\cof(c_1,...,c_a)$.

\item For the predicates that reference founded semantics, for each of the
  three truth values $t$, $P.t(c_1,...,c_a)$ is true in $I$ if
  $P(c_1,...,c_a)$ has truth value $t$ in $I$, and is false otherwise.

\item For the predicates that reference constraint semantics, $K_1\cocs(c)$
  is true in $I$ if $c$ is a model in the constraint semantics of $K_1$,
  and is false otherwise; the requirement that the CS-dependency relation
  is acyclic ensures that the constraint models of $K_1$ are determined
  before the semantics of $K_1\cocs(c)$ is considered.

\item If $c$ is a constraint model that provides a truth value for
  $P(c_1,...,c_a)$, then $c.P(c_1,...,c_a)$ has the same truth value in $I$
  that $P(c_1,...,c_a)$ has in $c$, otherwise it is undefined.
\end{itemize}
An interpretation $I$ of $K$ is {\em 2-valued} if every atom of $K$ is true
or false in $I$, that is, no atom is undefined.  Interpretations are
ordered by set inclusion $\subseteq$.

A {\em ground instance} of a rule $R$ is any rule that can be obtained from $R$ by expanding universal quantifications into conjunctions over all constants in the domain, instantiating existential quantifications with constants, and instantiating the remaining variables with constants.  

An interpretation is a {\em model} of a kunit if it contains all facts in
the kunit and satisfies all rules of the kunit (that is, for each ground
instance of each rule, if the body is true, then so is the conclusion),
when the rules are interpreted as formulas in 3-valued logic
\cite{fitting85}.  A collection of interpretations, one per kunit in a
program $\pgm$, is a {\em model} of $\pgm$ if each interpretation is a
model of the corresponding kunit.

The {\em one-step derivability} operator $T_K$ performs one step of inference using rules of $K$, starting from a given interpretation.  Formally, $C\in T_K(I)$ iff $C$ is a fact of $K$ or there is a ground instance $R$ of a rule in $K$ with conclusion $C$ such that the body of $R$ is true in $I$.

\mypar{Dependency graph}
The {\em dependency graph} $\dg(K)$ of kunit $K$ is a directed graph with a node for each predicate of $K$ that does not reference founded semantics and constraint semantics (including these predicates is unnecessary, because they cannot appear in conclusions), and an edge from $Q$ to $P$ labeled $\mathord{+}$ (respectively, $\mathord{-}$) if a rule whose conclusion contains $Q$ has a positive (respectively, negative) hypothesis that contains $P$.  If the node for predicate $P$ is in a cycle containing only positive edges, then $P$ has {\it circular positive dependency} in $K$; if it is in a cycle containing a negative edge, then $P$ has {\em circular negative dependency} in $K$.

\subsection{Founded semantics and constraint semantics of DA logic}
\label{subsec:semantics}

This subsection first defines founded semantics of DA logic without  meta-constraint \co{closed}, then extends the founded semantics to handle meta-constraint \co{closed}, and then defines the constraint semantics of DA logic.

\mypar{Founded semantics of DA logic without meta-constraint \co{closed}}
Intuitively, the {\em founded model} of a kunit $K$ without meta-constraint \co{closed}, denoted $\founded_0(K)$, is the least set of literals that are given as facts or can be inferred by repeated use of the rules. %
We define $\founded_0(K) = \lfpscc(\nameneg(\cmpl(K)))$, where functions $\cmpl$, $\nameneg$, and $\lfpscc$, are defined as follows.

\mypar{Completion}
The completion function, $\cmpl(K)$, returns the {\it completed} version of
$K$.  Formally, $\cmpl(K)=\addinv(\comb(K))$, where $\comb$ and $\addinv$
are defined as follows.

The function $\comb(K)$ returns the kunit obtained from $K$ by replacing
the facts and rules defining each uncertain and complete predicate $Q$ with a
single {\em combined rule} for $Q$,
defined as follows. (1) Transform the facts and rules defining $Q$ so they all have the same conclusion $Q(V_1, ..., V_a)$, by replacing each fact or rule $Q(X_1, ..., X_a) ~\IF~ B$ with $Q(V_1,...,V_a) ~\IF~ (\SOME~Y_1,...,Y_k ~|~$ $V_1=X_1 \land \cdots \land V_a=X_a \land B)$, where $V_1,...,V_a$ are fresh variables (i.e., not occurring in the given rules defining $Q$), and $Y_1,...,Y_k$ are all variables occurring in $X_1,...,X_a, B$, where $B$ denotes the entire body of the rule.  (2) Combine the resulting rules for $Q$ into a single rule defining $Q$ whose body is the disjunction of the bodies of those rules.  This combined rule for $Q$ is logically equivalent to the original facts and rules for $Q$.  This definition is the same as given for the core language in~\cite{LiuSto18Founded-LFCS,LiuSto20Founded-JLC}, except generalized to allow rule bodies that may contain disjunction.  Similar completion rules are used in Clark completion \cite{clark78} and Fitting semantics \cite{fitting85}.

The function $\addinv(K)$ returns the kunit obtained from $K$ by adding,
for each uncertain and complete predicate $Q$, a {\em completion rule} that
derives negative literals for $Q$.  The completion rule for $Q$ is obtained
from the inverse of the combined rule defining $Q$ (recall that the inverse
of $C~\IF~B$ is $\neg C~\IF~\neg B$), by putting the body of the rule in
negation normal form, that is, using equivalences of predicate logic to
move negation inwards and eliminate double negations, so that negation is
applied only to atoms.

\mypar{Least fixed point}

Explicit use of negation is eliminated before the least fixed point is computed, by applying the function $\nameneg$.  The function $\nameneg(K)$ returns the kunit obtained from $K$ by replacing each %
$\neg P(X_1, ..., X_a)$ with\lncsonly{\linebreak[5]} $P\cof(X_1, ..., X_a)$. 

The function $\lfpscc(K)$ uses a least fixed point to infer facts for each strongly connected component (SCC) in the dependency graph of $K$, as follows.  Let $S_1,...,S_n$ be a list of the SCCs in dependency order, so earlier SCCs do not depend on later ones; it is easy to show that any linearization of the dependency order leads to the same result for $\lfpscc$.  
For convenience, we overload $S_i$ to also denote the set of predicates in the SCC $S_i$.

Define $\lfpscc(K) = I_n$, where $I_0 = \emptyset$ and 
$I_i = \addneg(\lfp(T_{I_{i-1} \union \proj{K}{S_i}}), S_i)$
for $i \in 1..n$.  $\lfp(f)$ is the least fixed point of function $f$.  The least fixed point is well-defined, because 
$T_{I_{i-1} \union \proj{K}{S_i}}$ is monotonic, because the kunit $K$ was
transformed by $\nameneg$ and hence does not contain negation.  The
function $\addneg(I, S)$ returns the union of $I$ and the set of {\em
  completion facts} for predicates in $S$ that have meta-constraint \co{certain};
specifically, for each such predicate $P$, and for each combination of
values $c_1,...,c_a$ of arguments of $P$, if $I$ does not contain
$P(c_1,...,c_a)$, then $P\cof(c_1,...,c_a)$ is added as a completion fact.

\mypar{Founded semantics of DA logic with meta-constraint \co{closed}}

Informally, when a predicate of kunit $K$ has meta-constraint
\co{closed}, an atom $A$ of the predicate is false in an interpretation $I$,
called {\it self-false} in $I$, if every ground instance of rules that
concludes $A$, or recursively concludes some hypothesis of that rule
instance, has a hypothesis that is false or, recursively, is self-false in
$I$.

To formally define the set of self-false atoms, we first transform the rules of $K$ so that they do not contain disjunction, by putting the body of each rule $R$ containing disjunction into disjunctive normal form (DNF) and then replacing $R$ with multiple rules, one per disjunct of the DNF; this allows direct re-use of the following definitions of unfounded set and self-false atom from \cite{LiuSto18Founded-LFCS,LiuSto20Founded-JLC}, which do not take disjunction into account.

A set $U$ of atoms of kunit $K$ is an {\em unfounded set} of $K$ with respect to an interpretation $I$ of $K$ iff, for each atom $A$ in $U$, for each ground instance $R$ of a rule of $K$ with conclusion $A$, either (1) some hypothesis of $R$ is false in $I$ or (2) some positive hypothesis of $R$ for a closed predicate is in $U$; this is the usual definition of unfounded set \cite{van+91well}, except we inserted ``for a closed predicate''.  $\selffalse_K(I)$, the set of self-false atoms of kunit $K$ with respect to interpretation $I$, is the greatest unfounded set of $K$ with respect to $I$.
 
The founded semantics is defined by repeatedly computing the semantics
given by $\founded_0$ (the founded semantics without meta-constraint  \co{closed})
and then setting self-false atoms to false, until a least fixed point is
reached.  For a set $S$ of positive literals, let
$\neg \cdot S = \{ P\cof(c_1,...,c_a) \,|\, P(c_1,...,c_a) \in S\}$.  For a
kunit $K$ and an interpretation $I$, let $K \union I$ denote $K$ with the
literals in $I$ added to its body.  Formally, the founded semantics is
$\founded(K)=\lfp(F_K)$, where
$F_K(I) = \founded_0(K \union I) \union \neg \cdot \selffalse_K(\founded_0(K
\union I))$.

\mypar{Constraint semantics of DA logic}
Constraint semantics is a set of 2-valued models based on founded
semantics.  A {\em constraint model} of $K$ is a consistent 2-valued
interpretation $I$ of $K$ such that $I$ is a model of $\cmpl(K)$ and such
that $\founded(K) \subseteq I$ and
$\neg \cdot \selffalse_K(I) \subseteq I$.  Let $\constraint(K)$ denote the
set of constraint models of $K$.  Constraint models can be computed from
$\founded(K)$ by iterating over all assignments of true and false to atoms
that are undefined in $\founded(K)$, and checking which of the resulting
interpretations satisfy all rules in $\cmpl(K)$ and satisfy
$\neg \cdot \selffalse_K(I) \subseteq I$.

\subsection{Properties of DA logic semantics}

The following theorems express important properties of the semantics.

\begin{theorem}
  The founded model and constraint models of a program $\pgm$ are consistent.
  \label{thm:consistent}
\end{theorem}

\begin{myproof}
First we consider founded semantics.  Each kunit in the program defines a
distinct set of predicates, so consistency can be established one kunit at
a time. 
For each kunit $K$, the proof of consistency is a straightforward extension
of the proof of consistency of founded semantics \cite[Theorem
1]{LiuSto20Founded-JLC}.  The extension is to show that consistency holds
for the new predicates that reference founded semantics and constraint
semantics.

For predicates in \m{K} that reference founded semantics, we prove this for
each SCC $S_i$ in the dependency graph for $K$; the proof is by induction
on $i$.  The predicates used in SCC $S_i$ to reference founded semantics
have the same truth values as the referenced predicates in earlier SCCs.
These truth values are consistent because, by the induction hypothesis, the
interpretation computed for predicates in earlier SCCs is consistent.

For predicates in \m{K} that reference constraint semantics,
they have the same truth values as the referenced predicates in the
constraint models of the referenced kunits, and constraint models are
consistent by definition.

Next we consider constraint semantics.  Again note that constraint models
are consistent by definition.
\end{myproof}

\begin{theorem}
  The founded model of a kunit $K$ is a model of $K$ and $\cmpl(K)$.  The constraint models of $K$ are 2-valued models of $K$ and $\cmpl(K)$.
  \label{thm:model}
\end{theorem}

\begin{myproof} 
The proof that $\founded(K)$ is a model of $\cmpl(K)$ is essentially the same as the proof that $\founded(\pgm)$ is a model of $\cmpl(\pgm)$ \cite[Theorem 2]{LiuSto20Founded-JLC}, because the proof primarily depends on the behavior of $\cmpl$, $\addneg$, and the one-step derivability operator, and they handle atoms of predicates that reference founded semantics and constraint semantics in exactly the same way as other atoms.
Constraint models are 2-valued models of $\cmpl(K)$ by definition.
Any model of $\cmpl(K)$ is also a model of $K$, because $K$ is logically equivalent to the subset of
$\cmpl(K)$ obtained by removing the completion rules added by $\addinv$.
\end{myproof}

\begin{theorem}
  DA logic is decidable.
\end{theorem}

\begin{myproof}
DA logic has a finite number of constants from given facts, and has sets of finite nesting depths
bounded by the depths of CS-dependencies.  In particular, it has no function symbols to build infinite domains in recursive rules.  Thus, DA logic is over finite domains and is decidable.
\end{myproof}

Proving decidability of DA Logic is straightforward, but stating it explicitly is important, because DA logic supports recursion and allows nested constraint models to be used as constants.

\section{Additional examples}
\label{sec-examp}

We present additional examples that show the power of our language.
They are challenging or impossible to express and solve using prior 
languages and semantics.
For each example, we spell out the default meta-constraint for each predicate, in a topological-sort dependency order.
We use \co{-\,-} to prefix comments.

\subsection{Same different games}
The same win-not-win game can be over different kinds of moves, forming
different games, using kunit instantiation.  However, the fundamental
winning, losing, and draw situations stay the same, parameterized by the
moves.  The moves could also be defined easily using another kunit
instantiation.
\begin{myexample}
  Consider the following kunits.  First, \co{path\_unit} defines \co{path}
  recursively using \co{edge}: there is a path from \co{x} to \co{y} if
  there is a sequence of connected edges leading from \co{x} to \co{y}.
  Then, \co{win\_path\_unit} defines \co{link}, uses \co{path\_unit} to
  infer \co{path} with \co{edge} bound to \co{link}, and finally uses
  \co{win\_unit} in Section~\ref{sec-motiv} to determine winning, losing,
  and draw positions except with \co{move} bound to \co{path}.
  With default meta-constraints, \co{edge} and \co{path} are certain, \co{link} is
  certain, and \co{win} is complete.
\begin{code}
 kunit path_unit:
   path(x,y) \IF edge(x,y)
   path(x,y) \IF edge(x,z) \AND path(z,y)
   
 kunit win_path_unit:
   link = \{(1,2), (1,3), ...\}   -- shorthand for link(1,2), link(1,3), ...
   use path_unit (edge = link)  -- instantiate path_unit with edge replaced\lncsonly{} by link
   use win_unit (move = path)   -- instantiate win_unit with move replaced\lncsonly{} by path
\end{code}
Alternatively, in \co{win\_path\_unit}, one could define \co{edge} instead
of \co{link}, and then use \co{path\_unit} without replacing the name
\co{edge} to \co{link}, as follows.
\begin{code}
 kunit win_path_unit:
   edge = \{(1,2), (1,3), ...\}   -- define edge in place of link
   use path_unit ()             -- use path_unit without replacing edge to\lncsonly{} link
   use win_unit (move = path)
\end{code}\Vex{-3}
\end{myexample}

\subsection{Defined from undefined positions}
Sets and predicates can be defined using the set of values of arguments for
which a given predicate is undefined.  This is not possible in previous
3-valued logic like WFS, because anything depending on undefined can only
be undefined.

\begin{myexample} 
  Consider the following \co{draw\_unit}.  It defines \co{move} and uses
  \co{win\_unit}.  Then, using the result of win-not-win game,
  predicate \co{move\_to\_draw} defines the set of positions that have a
  move to a draw position, and predicate \co{reach\_from\_draw} defines the
  set of positions that are reachable by following a path of special moves
  from a draw position.
  With default meta-constraints, \co{move} is certain, 
  \co{win} is complete, and \co{move\_to\_draw},
  \co{special\_move}, \co{path}, and \co{reach\_from\_draw} are certain.

\begin{code}
 kunit draw_unit:
   move = \{(1,1), (2,3), (3,1)\}   
   use win_unit ()\medskip
   move_to_draw(x) \IF move(x,y) \AND win.U(y)\medskip
   special_move = \{(1,4), (4,2)\}
   use path_unit (edge = special_move)\medskip
   reach_from_draw(y) \IF  win.U(x) \AND path(x,y)
\end{code}
In \co{draw\_unit}, we have \co{win.U(1)}, that is, 1 is a draw position.
Then we have \co{move\_to\_draw(3)} to be true, and we have
\co{reach\_from\_draw(4)} and \co{reach\_from\_draw(2)} to be true.

Note that we could copy the single win rule here in place of \co{use
  win\_unit ()} and obtain an equivalent \co{draw\_unit}.  We avoid copying
when possible because this is a good principle, and in general, a kunit may
contain many rules and facts.
\end{myexample}

\subsection{Unique undefined positions}
Among the most critical information is 
assertions that have a unique true or false value in all possible
ways of satisfying given constraints but cannot be determined to
be true by just following founded reasoning.  Having both founded semantics
and constraint semantics at the same time allows one to find such
information.

\begin{myexample}
  Consider the following two kunits.  
  First, 
  \co{pa\_unit} defines \co{prolog}, \co{asp}, and \co{move} and uses
  \co{win\_unit}.
  Then, \co{cmp\_unit} uses \co{pa\_unit} and defines \co{unique(x)} to be
  true if (1) \co{win(x)} is undefined in founded semantics, (2) a
  constraint model of \co{pa\_unit} exists, and (3) \co{win(x)} is true in
  all models in the constraint semantics.
  With default meta-constraints, predicates \co{prolog} and \co{asp} are complete, \co{move} is specified to be closed, 
  \co{win} is complete,
  and predicate \co{unique} is certain. %
  Note that \co{prolog}, \co{asp}, and \co{move} cannot be certain because
  \co{prolog} and \co{asp} are defined with negation in recursion, and
  \co{move} depends on \co{prolog} and \co{asp}.\m{\!\!}
\begin{code}
 kunit pa_unit:
   prolog \IF \NOT asp
   asp \IF \NOT prolog
   move(1,0) \IF prolog
   move(1,0) \IF asp
   closed(move)
   use win_unit ()\medskip
 kunit cmp_unit:
   use pa_unit ()\medskip
   unique(x) \IF win.U(x) \AND \SOME m \IN pa_unit.CS\lncsonly{} \AND \EACH m \IN pa_unit.CS | m.win(x)
\end{code}
In \co{pa\_unit}, founded semantics gives \co{move.U(1,0)} (because
\co{prolog} and \co{asp} are undefined),
\co{win.F(0)} (because there is no move from \co{0}), and \co{win.U(1)}
(because \co{win(1)} cannot be true or false).

Constraint semantics \co{pa\_unit.CS} has two models: 
\co{\{prolog, move(1,0), %
win(1)\}} and\linebreak \co{\{asp, move(1,0), %
win(1)\}}.  We see that \co{win(1)} is true in all two models. 
So \co{win.U(1)} from founded semantics is imprecise.  

In \co{cmp\_unit}, \co{unique(1)} is true.  That is,
\co{win(1)} is undefined in founded semantics, a constraint model 
exists, and \co{win(1)} is true in all models in the constraint
semantics.
\end{myexample}

\subsection{Multiple uncertain worlds}
Given multiple worlds each corresponding to a different model, different uncertainties
can arise from different worlds,
yielding multiple uncertain worlds.
It is simple to represent this using predicates that are possibly 3-valued
and that are parameterized by a 2-valued model.

\begin{myexample}
  Consider the following two kunits.  The game in \co{win\_unit2} uses
  \co{win\_unit} on a set of moves.  The game in \co{win\_set\_unit} has
  its own moves, but a move is valid if and only if it starts from a
  position that is a winning position in a model in the constraint
  semantics of \co{win\_unit2}.
  With default meta-constraints, \co{move} in both kunits are certain,
  \co{win} is complete,
  \co{valid\_move} is certain, \co{valid\_win} is complete, and  \co{win\_some} and \co{win\_each} are certain.

\begin{code}
 kunit win_unit2:
   move = \{(1,4),(4,1)\}
   use win_unit ()\medskip
 kunit win_set_unit:
   move = \{(1,2),(2,3),(3,1),(4,4),(5,6)\}
   valid_move(x,y,m) \IF move(x,y), win_unit2.CS(m), m.win(x)\medskip
   use win_unit (move = valid_move(m), win = valid_win(m))\medskip
   win_some(x) \IF valid_win.T(x,m)
   win_each(x) \IF win_some(x) \AND \EACH m \IN win_unit2.CS | valid_win.T(x,m)
\end{code}
In \co{win\_unit2}, there is a 2-move cycle.
The constraint semantics \co{win\_unit2.CS} is a set of two models, 
say \co{\{m1,m2\}}, where \co{m1.win = \{1\}} and \co{m2.win = \{4\}}.
That is, in \co{m1}, position \co{1} is winning, and position \co{4} is
not, and in \co{m2}, the situation is the opposite.

In \co{win\_set\_unit}, each model \co{m} in \co{win\_unit2.CS} leads to a
separately defined \co{valid\_move} under argument \co{m}.
In \co{m1}, only \co{move(1,2)} starts from the winning position \co{1},
and in \co{m2}, only \co{move(4,4)} starts from the winning position
\co{4}.  %
So \co{valid\_move} is true for only \co{valid\_move(1,2,m1)} and
\co{valid\_move(4,4,m2)}.

The separate \co{valid\_move} under argument \co{m} is then used to define
a separate \co{valid\_win} under argument \co{m}, by instantiating
\co{win\_unit} with predicates \co{move} and \co{win} bound to
\co{valid\_move} and \co{valid\_win}, respectively, with the additional
argument \co{m}.
This yields \co{valid\_win} being true for only \co{valid\_win(1,m1)}.

Finally, \co{win\_some(x)} is true for any position \co{x} such that
\co{valid\_win(x,m)} is true for some model \co{m}, and \co{win\_each(x)}
is true if \co{win\_some(x)} is true and \co{valid\_win(x,m)} is true for
all models \co{m} in \co{win\_unit2.CS}.
The result is that \co{win\_some} is true for only \co{win\_some(1)} and is
false for all other positions, and \co{win\_each} is false for all
positions.
\end{myexample}

\section{Restricted parameters and circular uses of kunits}
\label{sec-disc}

Knowledge units are similar to modules in that they provide a means to
organize the knowledge expressed as logic rules and constraints.  We
discuss extensions that allow knowledge units to have specially specified
parameters, and have circular uses.  We describe the pros and cons of supporting them and show that there is no real difficulty in supporting them.

\subsection{Units with restricted parameters}
Knowledge units as described in Section~\ref{sec-lang} do not need
specially specified parameters.  Any predicate in a kunit is in fact a
parameter that can be instantiated with any predicate of the same number of
arguments, or even with additional arguments if desired.

Some people may be accustomed to using modules or components with a
specially specified set of parameters, where all uses of the module or
component must instantiate exactly this restricted set of parameters.  This
is straightforward to add to DA logic, by simply specifying some of the
predicates in a kunit as this restricted set of parameters of the kunit.

There are both pros and cons with specially specified parameters.
\begin{itemize}
\item The advantage is that one can hide the remaining predicates of the
  kunit from uses of the kunit.  Changes to the hidden predicates will not
  affect uses of the kunit so long as the changes do not affect the
  specially specified parameters.

\item The disadvantage is that if a hidden predicate becomes useful outside
  the kunit, the predicate must be added to the specially specified
  parameters to be used.  Furthermore, this change is not limited to new
  uses of this kunit, but requires changes to all previous uses of the
  kunit.
\end{itemize}

Knowledge units with no restriction on parameters are more general and
powerful for knowledge representation, for at least two reasons.
\begin{enumerate}
\item They can be used in any way that is easy and clear, with any
  combination of instantiated predicates that is needed, without changing
  the kunit or any previous uses of the kunit.

\item They encourage all predicates in a kunit to be carefully defined for
  clarity and reusability, eliminating the need to hide predicates that are
  not externally used.
\end{enumerate}
Procedural programming benefits greatly from hiding internal details,
because additional variables and parameters are most often used for
efficiency reasons.  In DA Logic, rules are declarative specifications, and
hiding such specifications is generally unnecessary.

Nevertheless, to support hiding certain predicates, a kunit can specially
specify which predicates can be externally used, as follows:
\begin{code}
 kunit \p{K} (\p{preds}):
   \p{body}
\end{code}
where \p{preds} is a set of predicates in \p{K} that can be instantiated or
can be used outside \p{K}.
The \co{use} clause 
\[
\co{use}~K~(P_1=Q_1(Y_{1,1}, ..., Y_{1,b_1}),\, ...,\, P_n=Q_n(Y_{n,1}, ..., Y_{n,b_n}))
\]
does not need any change.  The semantics is extended to check that each
predicate \p{P_i} is in the set \p{preds} of specially specified predicates
of kunit \p{K}, and to ensure that there is no external use of predicates
not in \p{preds}.

Note that this extension still allows 
each use of a kunit to instantiate any subset of the specially specified
predicates of the kunit.  This design is more general than parameterized
module systems in which each module has a fixed set of parameters, all of
which must be instantiated at every use.

\subsection{Units with circular uses}
Uses of knowledge units as described in Section~\ref{sec-lang} must form
acyclic dependencies.  

Some people may be accustomed to module systems that allow circular
uses of modules. %
Allowing circular uses has both pros and cons.
\begin{itemize}
\item The advantage is that modules could be smaller and more flexible, 
  and could use one another recursively.

\item The disadvantage is that the dependencies between predicates in  
  modules with circular uses may be difficult to determine and understand.
\end{itemize}

Knowledge units with no circular uses are easier to understand, for at
least two reasons.
\begin{enumerate}
\item Dependencies between predicates defined in the knowledge units are
  clearer at a high level, because they must follow the tree of
  dependencies between kunits.  With circular uses of kunits, all
  predicates defined in those kunits potentially depend on each other,
  depending on the details of their definitions.

\item Within a kunit, predicates easily capture any structure including
  cyclic graphs and the trivial case of recursive structures like trees,
  and recursive rules can easily define mutually dependent predicates.
\end{enumerate}

Nevertheless, to support circular uses of kunits in DA logic, we can
eliminate the requirement that the use-dependency relation is acyclic, and
extend the semantics of \co{use} to handle circularity
as follows.  Recall from Section \ref{sec-lang} that using a kunit has the
effect of instantiating the body using the specified substitution and then
inlining the result at the use.  To support circular uses, the algorithm is
extended to keep track of which uses of kunits have already been
instantiated and inlined.  The effect of using a kunit is to check whether
the same use of the kunit has already been instantiated and inlined, and if
so, do nothing, otherwise instantiate and inline it.

DA logic with this extension is still decidable, because there is only a
finite number of possible uses of kunits in a program.

\section{Related work and conclusion}
\label{sec-related}

Many logic languages and semantics have been proposed. Several overview 
articles~\cite{apt1994negation,prz94well,ramUll95survey,fitting2002fixpoint,trusz18sem-wbook}
give a good sense of the complications and challenges when there is
unrestricted negation in recursion.
Notable different semantics include Clark completion~\cite{clark78} 
and similar additions, e.g.,~\cite{lloyd84making,sato84transformational,Jaffar+86some,chan88constructive,foo88deduced,stuckey91constructive}, 
Fitting semantics or Kripke-Kleene semantics~\cite{fitting85},
supported model semantics~\cite{apt88},
stratified semantics~\cite{van86negation,apt88},
WFS~\cite{van88unfounded,van+91well}, %
and SMS~\cite{GelLif88}.
Note that these semantics disagree, in contrast to different styles of semantics that agree~\cite{ershov1987semantic}.

There are also a variety of works on relating and unifying different semantics.
These include Dung's study of relationships~\cite{dung1992relations}, 
partial stable models, also called
stationary models~\cite{prz94well},
Loop formulas~\cite{lin2004assat},
FO(ID)%
~\cite{denecker2008logic},
and founded semantics and constraint semantics~\cite{LiuSto18Founded-LFCS,LiuSto20Founded-JLC}.
FO(ID) is more powerful than works prior to it, by supporting both first-order logic and inductive definitions while also being similar to SMS~\cite{bruynooghe2016first}.  However, it does not support any 3-valued semantics.  Founded semantics and constraint semantics uniquely unify different semantics, by capturing their different assumptions using predicates declared to be certain, complete, and closed, or not.

However, founded semantics and constraint semantics by themselves do not
provide a way for different semantics to be used for solving different
parts of a problem or even the same part of the problem.  DA logic supports
these, and supports everything completely declaratively, in a unified
language.

Specifically, DA logic allows different assumptions under different
semantics to be specified easily as meta-constraints, and allows the
results of different semantics to be built upon, including defining
predicates using
atoms that have truth value undefined in a 3-valued model and using models
in a set of 2-valued models, and parameterizing predicates by a set of
2-valued models.  More fundamentally, DA logic allows different parts of a
problem to be solved with different knowledge units, where every predicate
is a parameter that can be instantiated with new predicates, including new
predicates with additional arguments.  These are not supported in prior
languages.

Among many directions for future work, one particularly important and
intriguing problem is to study optimal algorithms and precise complexity
guarantees, similar to~\cite{LiuSto09Rules-TOPLAS}, for inference and
queries for DA logic.

\lncsonly{}

\mypar{Acknowledgments.} 
We thank the anonymous referees of this journal article and our LFCS 2020 paper \cite{LiuSto20LogicalConstraints-LFCS} for their thoughtful reviews and detailed and helpful comments.

\articleonly{\bibliographystyle{alpha}}
\lncsonly{}
\def\bibdir{../../../bib}                        %
\bibliography{\bibdir/strings,references,\bibdir/liu,\bibdir/math}
\end{document}